\RequirePackage{amsmath}
\documentclass[structabstract]{aa}

\usepackage[utf8]{inputenc}

\usepackage{amsmath}

\usepackage{calc}

\usepackage{longtable}
\usepackage{booktabs}
\usepackage{array}

\usepackage{color}

\usepackage[pdftex]{graphicx}
\usepackage{subfigure}

\usepackage{txfonts}
\usepackage{ae}

\usepackage{natbib}
\bibpunct{(}{)}{;}{a}{}{,} 

\usepackage{units}

\usepackage{textcomp}
\usepackage{gensymb}

\usepackage[colorlinks=true,linkcolor=blue,citecolor=blue,bookmarks]{hyperref}

\begin{document}

\title{Morphometric analysis in gamma-ray astronomy\\
  using Minkowski functionals:}
\subtitle{II. Joint structure quantification}

\author{M.~A.~Klatt\inst{\ref{kit},\ref{theo1},\ref{ecap}}
  \thanks{\email{michael.klatt@kit.edu}}
  \and
  K.~Mecke\inst{\ref{theo1},\ref{ecap}}
}

\titlerunning{Morphometric analysis in gamma-ray astronomy}
\authorrunning{M.~A.~Klatt \and K. Mecke}

\institute{
  Karlsruhe Institute of Technology (KIT), Institute of Stochastics, Englerstr. 2, 76131 Karlsruhe, Germany\label{kit}
  \and
  Institut für Theoretische Physik,
  Universität Erlangen-Nürnberg,
  Staudtstr. 7,
  91058 Erlangen, Germany\label{theo1}
  \and
  Erlangen Centre for Astroparticle Physics,
  Universität Erlangen-Nürnberg,
  Erwin-Rommel-Str. 1,
  91058 Erlangen, Germany\label{ecap}
}

\date{Received \dots / Accepted \dots}


\abstract
{} 
{ We pursue a novel morphometric analysis to detect sources in very-high-energy gamma-ray counts maps by structural deviations from the background noise.
}
{ Because the Minkowski functionals from integral geometry quantify the shape of the counts map itself, the morphometric analysis includes unbiased structure information without prior knowledge about the source.
  Their distribution provides access to intricate geometric information about the background.
  We combine techniques from stochastic geometry and statistical physics to determine the joint distribution of all Minkowski functionals.
}
{ We achieve an accurate characterization of the background structure for large scan windows (with up to $15\times15$ pixels), where the number of microstates varies over up to 64 orders of magnitude.
  Moreover, in a detailed simulation study, we confirm the statistical significance of features in the background noise and discuss how to correct for trial effects.
  We also present a local correction of detector effects that can considerably enhance the sensitivity of the analysis.
}
{ In the third paper of this series, we will use the here derived refined structure characterization for a more sensitive data analysis that can detect formerly undetected sources.
}

\keywords{
  Methods: data analysis --
  Methods: statistical --
  Techniques: image processing --
  Gamma rays: diffuse background
}

\maketitle


\section{Morphometric source detection in gamma-ray astronomy}
\label{Introduction}

The unavoidable background noise in ground-based Very-High Energy (VHE) gamma-ray astronomy exhibits a rich and complex structure.
Although a source signal might be weak, some sources could still be detected if the complex structure of the background noise was better understood.
A quantification of the shape of the counts map itself can extract additional information beyond the simple total number of counts without any assumption about potential sources in the field of view.
In the first paper of this series, we introduced a novel approach to data analysis in VHE gamma-ray astronomy, where we use this geometric information to detect sources~\citep{KlattEtAl2012, GoeringKlattEtAl2013}.

The commonly applied hypothesis test by \citet{LiMa1983} is only based on the excess of gamma-ray counts on top of the expected background.
Such techniques based only on the total count of events, which omit all geometric information that could help to detect sources, were designed to study point sources.
However an increasing number of large extended sources and even diffuse VHE emissions are observed~\cite{rxj1713, velajr, galcen}.
There other powerful techniques that include additional information are full likelihood fits of models to the measured data, for example, used by high-energy gamma-ray telescopes like EGRET~\citep{egretlike} or Fermi/LAT~\citep{fermilat}.
However, the outcome of the fit will strongly depend on the model and on the a-priori knowledge about the sources.

In contrast to this, our morphometric analysis is a null hypothesis test that only depends on the background model.
Morphometric valuations from integral geometry, the so-called Minkowski functionals~\citep{SchroederTurketal:2010jom, SchroederTurketal2010AdvMater}, allow us to efficiently and comprehensively quantify the shape of the counts map itself.
Thus we are looking for any statistically significant structural deviation from the background noise, and there is no need for a priori modeling of potential sources.
In this sense, the Minkowski functionals serve as unbiased detectors of source structure or in general of inhomogeneities in background noise.
They can extract the essential features and thus formerly undetected sources can eventually be detected by a refined structure characterization.

In astronomy, the Minkowski functionals are already successfully applied to
study point processes in cosmology and the large-scale structure of the
universe~\citep{MeckeBuchertWagner1994, Colombi2000, KerscherEtAl2001, KerscherMecke2001, wiegand_direct_2014},
to analyze exotic states of nuclear matter in supernova explosions~\citep{SchuetrumpfKlattEtAl2013, SchuetrumpfKlattEtAl2014},
and to search for non-Gaussianity in the cosmic microwave background~\citep{WinitzkiKosowsky1998,Schmalzing1999, CurtoEtAl2008, Gay2012, Ducout2013}.

In the first paper of this series, we presented our morphometric analysis to gamma-ray astronomy.
Therefore, we first explained the structure characterization of gamma-ray sky maps with Minkowski functionals and then defined a null hypothesis test that determines the statistical significance of structural deviations.
Minkowski sky maps detect local structural features.
However, in the first paper we only used single functionals to analyze small windows, which limited the geometric information that could actually be included in the analysis.

The simultaenous characterization of the background structure of large observation windows with all Minkowski functionals provides access to much more intricate geometric information.
However, such a detailed knowledge about the shape of the noise also requires advanced techniques for accurate estimations of the structure distributions.
This crucial step is here achieved.
We derive precise estimates of the joint probability distribution of the Minkowski functionals.
The calculation is based on number of possible configurations which vary from $\mathcal{O}(10^{1})$ to $\mathcal{O}(10^{64})$ for a $15\times 15$ b/w image.

In the third paper of this series, we will use this refined structure characterization to demonstrate an increase in sensitivty due to the additional geometric information.

In Sec.~\ref{sec_gamma_short_overview}, we shortly summarize the most important steps of the morphometric analysis.
In Sec.~\ref{sec_gamma_structure_probab_distr}, we derive accurate estimates of the joint distribution of Minkowski functionals, which we achieve by combining analytic knowledge of the structure distributions with an efficient algorithm from statistical physics.
This main result of the article allows us to analyze larger scan windows up to $15\times 15$ pixels that contains complex structural information in contrast to the so far accessible window sizes.

In Sec.~\ref{sec_gamma_trial_factor}, we provide a conservative yet tight estimate of the trial factor, which is needed because of the repeated hypothesis tests at different thresholds.
Moreover, we show that the joint characterization of the structure using all three Minkowski functionals does not cause any distinct changes in the trial factor compared to a simple analysis based only on a single functional.
This allows to compare in the third paper the statistical significance of structural deviations from the background noise that are quantified either by simply the area or by all three Minkowski functionals.
In Sec.~\ref{sec_conclusion}, we summarize our results and make concluding remarks.

In Appendix~\ref{sec_gamma_detector_acceptance}, we present a technique to subtract known point-like sources from the observations as well as an optimal correction of variations in the detector acceptance with a minimum supression of source signals.
We demonstrate that this improvement, which is especially relevant for the application to real data, can lead to a significant increase in sensitivity of the morphometric analysis\footnote{Parts of this article are from the PhD thesis of one of the authors~\citep{Klatt2016}.}.


\section{The statistical significance of structural deviations}
\label{sec_gamma_short_overview}

The structure characterization via Minkowski functionals is explained in detail in the first paper of the series.
There we also rigorously defined the null hypothesis test.
Here we only shortly summarize the most important steps.

\subsection{The shape of counts maps}

The Minkowski funtionals allow for a sensitive and comprehensive structure characterization of black-and-white images~\citep{MantzJacobsMecke2008, SchroederTurketal2010AdvMater, SchroederTurkEtAl2013NewJPhys}.
They are efficient shape descriptors from integral geometry that quantify in arbitrary dimensions all motion invariant, continuous and additive shape information of convex bodies~\citep{Hadwiger1957, SchneiderWeil2008}.
A functional is additive, if its value for two disjoint bodies is given by the union of the single functional values; for example, the perimeter of two disjoint clusters of black pixels is given by the sum of the two single perimeters.
The additivity property implies both robustness against noise and efficiency (with a computation time that scales linearly with the system size).

For a two dimensional black-and-white pixelated image, these Minkowski functionals are given by three intuitive quantities, namely,
the area $A$, perimeter $P$, and Euler characteristic $\chi$.
The latter is a topological constant; it is the number of clusters minus the number of holes.

To quantify the shape of the counts map, it is turned into a series of black-and-white images via thresholding.
For each threshold $\rho$, a pixel is set to white if the corresponding number of counts is smaller than the threshold, otherwise it is set to black.
The Minkowski functionals then quantify the shape of these black-and-white images.

\subsection{Global null hypothesis test}
\label{sec_gamma_global_null_hypothesis_test}

Are at a given threshold the measured values of the Minkowski functionals compatible with the shape distribution of the background noise?
To answer this question, a null hypothesis test has to be defined to detect significant structural deviations from the expected background structure.
First, we have to define a model for the background noise.
Then the null hypothesis is that there are only background signals.

In ground-based VHE gamma-ray astronomy, most background events are caused by VHE hadrons.
Because these hadrons loose direction correlations in interstellar magnetic fields, they arrive at our atmosphere as a uniform flux from every direction.
Therefore, we can well model the background noise by a so-called Poisson point process (or ``Complete Spatial Randomness''), that is, we assume random, completely independent, homogeneously distributed background events.
For the binned counts map, this results in an independent number of counts in each (equal area) bin, which follow a Poisson distribution.
The background intensity, which is the mean number of counts per bin, is here denoted by $\lambda$.

In a real measurment, detector effects and nonuniform exposure of the sky distort the homogeneous and isotropic background.
However, as we have demonstrated in the first paper of this series, our corrections of these effects allow for an analysis of real data.
In the Appendix~\ref{sec_gamma_local_detector}, we present an improved correction of such detector effects that can considerably enhance the sensitivity of the analysis.

Under the assumption of the null hypothesis (that there are only background signals) the probability distribution $\mathcal{P}$ of the Minkowski functionals is well defined.
In contrast to the first paper, we here do not only consider the probability distributions of single functionals but the joint distribution of all Minkowski functionals.
The joint probability distribution $\mathcal{P}(A,P,\chi)$ of the area $A$, perimeter $P$, and Euler characteristics $\chi$ of the black-and-white image clearly depends on the threshold.
As mentioned above, it encodes an advanced knowledge about the structure of the background noise.
It must once be determined with high accuracy so that we are able to perform a null hypothesis test.
The methods to derive the probability distribution are explained in detail in Sec.~\ref{sec_gamma_structure_probab_distr}.

The central idea of the morphometric analysis is to detect gamma-ray sources (or more precisely to reject the null hypothesis) by identifying structures with a statistically significant deviation from the expected behavior of background noise.
Following a scheme by \citet{NeymanPearson1933} for constructing a most efficient hypothesis test with no constraints on alternative hypotheses, we define the \textit{compatibility} $\mathcal{C}$ of a measured triplet $(A,P,\chi)$ with the null hypothesis:
\begin{align}
  \mathcal{C}(A,P,\chi) &= \sum_{\mathcal{P}(A_i,P_i,\chi_i) \le \mathcal{P}(A,P,\chi)} \mathcal{P}(A_i,P_i,\chi_i)
  \label{eq:compatibility}
\end{align}
It is the probability for the appearance of a structure that is even less likely than the measured structure. 

We reject the null hypothesis of a pure background measurement if the compatibility is lower than $0.6\cdot10^{-6}$.
This hypothesis criterion corresponds to the commonly used $\unit[5]{\sigma}$ deviation in the sense that a normally distributed random variable deviates from the expected value by at least $\unit[5]{\sigma}$ with a probability of approximately $0.6\cdot10^{-6}$.

It is more convenient to avoid small probabilites and to define a measure that gets larger if the structural deviation is stronger.
We therefore define the \textit{deviation strength} $\mathcal{D}$ as the logarithm of this likelihood value:
\begin{align}
  \mathcal{D}(A,P,\chi) :=& - \log_{10}\mathcal{C}(A,P,\chi) \, .
  \label{eq:dev_strength}
\end{align}
The null hypothesis criterion is rejected if the deviation strength is larger than $6.2$.

\subsection{Minkowski sky maps}

To detect local structural deviations and visualize the sources, we have defined Minkowski sky maps.
The analysis is restricted to a small scan (or sliding) window.
We assign to the central pixel the maximum value of the deviation strength over all thresholds.
These repeated trials for different thresholds must be taken into account.
We define a trial correction and perform a simulation study in Sec.~\ref{sec_gamma_trial_factor}.
An iteration of the scan-window over the whole counts map yields the Minkowski sky map, similar to a significance map constructed using the approach by \citet{LiMa1983}.
It shows the statistical significance of local features in different regions of the field of view.


\section{Joint structure characterization}
\label{sec_gamma_structure_probab_distr}

The most curcial and at the same time most difficult step in the preparation of the morphometric analysis is to determine the joint probability distribution of the Minkowski functionals.
It is via the joint probability distribution, that we include the accurate shape information about the background noise in our morphometric analysis.
Often in stochastic geometry, only first and second moments of distributions are determined because analytic calculations or precise numerical estimates of probability distribution form complex problems~\citep{Adler1981, Stoyan1987}.
Here we solve this problem here by using advanced techniques from statistical physics, which have been designed to study phase transitions.

\subsection{Structure probability distribution and density of states}

For the homogeneous Poisson field, the question for the probability distribution can be reformulated as a question for the ``Density of States'' (DoS) $\Omega$.
The latter is defined as the number of different b/w images (``microstates'') with the same Minkowski functionals (``macrostate'').
The probability $\mathcal{P}(A,P,\chi)$ to find a configuration with area $A$, perimeter $P$ and Euler characteristic $\chi$ is then given by
\begin{align}
  \mathcal{P}(A,P,\chi) = \Omega(A,P,\chi) \cdot p_{\rho}^A \cdot
  (1-p_{\rho})^{N^2-A},
  \label{eq_gamma_PandDos}
\end{align}
with $p_{\rho} = \sum_{i=\rho}^{\infty}{\frac{\lambda^i}{i!} e^{-\lambda}}$ being the probability that a pixel is black.
In contrast to the probability distribution, the density of states is independent of the probability $p_{\rho}$ that a pixel is black, i.e., of the intensity $\lambda$ and the threshold $\rho$.
Once the density of states is determined, the probability distribution can easily be calculated for any observed values of the Minkowski functionals.

For small window sizes up to $6\times 6$, this problem can be solved by
simply computing the Minkowski functionals of all possible b/w
images. However, the total number of microstates increases
super-exponentially $\sum_{A,P,\chi} \Omega = 2^{N^2}$; for $6\times
6$ the total number of states is already $\sum_{A,P,\chi} \Omega =
68.719.476.736$.

Using combinatorial techniques from the theory of partition numbers, an
algorithm can be constructed whose complexity only increases
sub-exponentially in the number of bins~\citep{Goering2012}. However,
the need of memory very quickly outgrows today's available hardware.
The density of states can be determined only for up to a $7\times 7$
observation window.

However, it is crucial to use larger observation windows, because only
those can incorporate considerable structural information leading to a
significantly more sensitive hypothesis test. 
Already for small sliding windows, there is a slight increase in
sensitivity if the source has strong structural features within the
sliding window, i.e., a strong intensity gradient like a Gaussian peak
with an extension smaller than the window. However, larger window sizes are needed for application to real data.

The DoS for larger systems cannot be determined analytically, but only
numerically. However, a simple Monte Carlo simulation, which estimates
the DoS of a macrostate by the frequency of its appearance in a simple
sampling of the space of microstates, is not sufficient. Although a
configuration might be entropically suppressed in the simple sampling,
i.e., an unlikely configuration of the excursion set of a Poisson
background noise, it might be likely to appear in the presence of a
source.
If such a yet unknown macrostate is detected, it will result in an
infinite deviation strength because the numerical estimate of the
compatibility is zero. The missing configurations thus produce
unacceptable artifacts in the Minkowski sky map and the null
hypothesis is no longer well-defined.

An intelligent algorithm is needed, which is able to give, e.g., for a
$15\times 15$ observation window a reliable estimate for both
macrostates with a DoS $\mathcal{O}(10^{64})$ and for those with a DoS
of $\mathcal{O}(1)$. We therefore apply the so-called
Wang-Landau algorith developed in condensed matter physics for
studying phase transitions and critical
phenomena~\citep{WangLandau2001PRE, WangLandau2001PRL}.

It uses a non-Markovian random walk to efficiently sample the DoS; by
continuously adjusting the density of states a locally flat histogram
is achieved. More precisely, it samples the space of microstates
according to a probability distribution that is inverse to the
DoS. Thus, each macrostate is encountered equally often, which is
called flat histogram sampling. The a priori unknown DoS is estimated
by gradually improving an initial estimate, which is why the algorithm
is non-Markovian. The initial estimate can, for example, be a constant
for all macrostates, which corresponds to no initial information.

Using the Wang-Landau algorithm allows to
determine the joint probability distribution of the Minkowski
functionals for a Poisson field up to a $15\times 15$ observation
window, with a high precision and well suited for accurate
hypothesis tests in gamma-ray astronomy.

\subsection{Wang-Landau algorithm combined with analytic DoS}

The Wang-Landau algorithm uses a random walk through the space of 
macrostates, the so-called energy space, where the microstates
are, e.g., black and white images. The energy space is scanned by
randomly changing the microstate, e.g., randomly changing the pixels
from black to white and vice versa~\citep{LandauEtAl2004,
  TsaiEtAl2006}. In each step, the numerical estimate of the density
of states $\hat{\Omega}$ is adjusted: if the resulting macrostate is
$(A,P,\chi)$, then $\hat{\Omega}(A,P,\chi)\rightarrow
f\cdot\hat{\Omega}(A,P,\chi)$. A finite modification factor $f$
introduces a systematic error and has to decrease during the
simulation. Steps to macrostates with a smaller density of states are
always accepted, but changes which result in a macrostate with a
larger density of states are only accepted with a probability that is
proportional to the ratio of the density of states:
$$
\text{Prob}_{\text{accept}}\left[\left(\begin{array}{c}
A_1\\ P_1\\ \chi_1
\end{array}\right)\rightarrow\left(\begin{array}{c}
A_2\\ P_2\\ \chi_2
\end{array}\right)\right] = \min\left\{1,\frac{\hat{\Omega}\left(A_1,P_1,\chi_1\right)}{\hat{\Omega}\left(A_2,P_2,\chi_2\right)}\right\}\;.
$$
Thus, a flat histogram in the energy space, i.e., the number of
visits of macrostate, is achieved.  If the histogram is sufficiently
flat, e.g., if the minimum value is at least $80\,\%$ of the average,
the modification factor is replaced by its square root.

\begin{figure}
  \centering
  \includegraphics[width=0.46\linewidth]{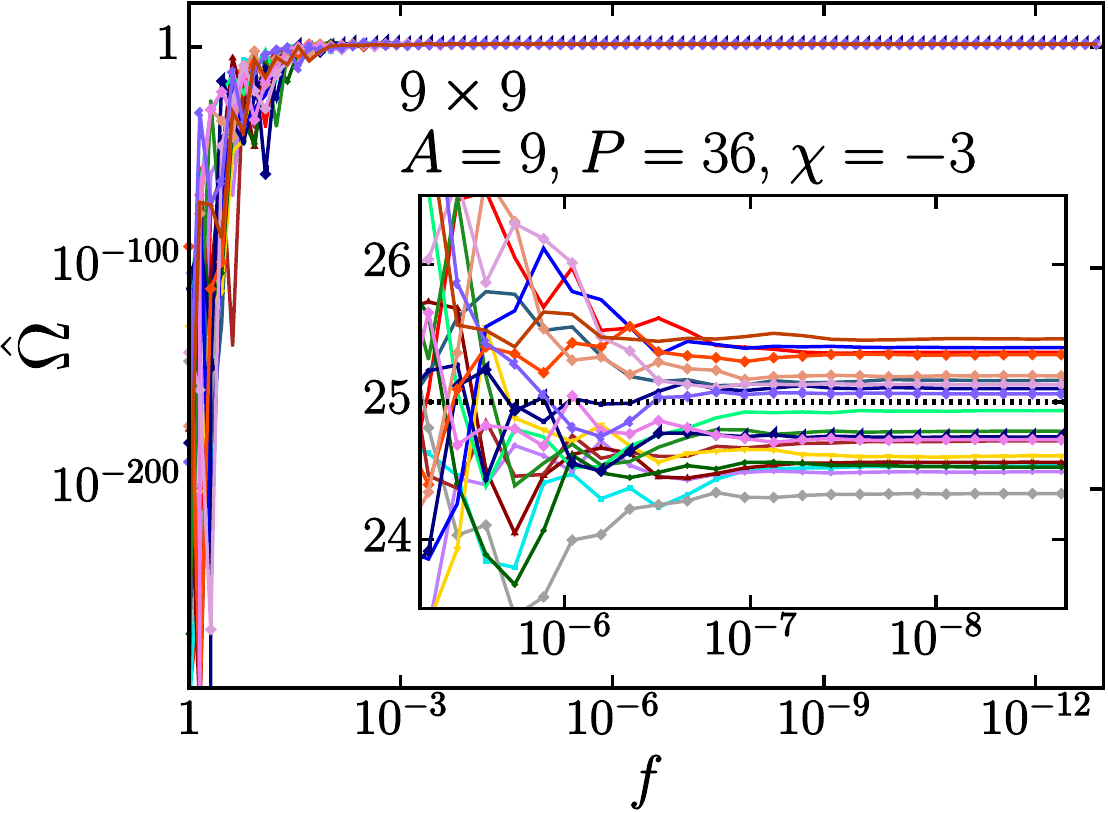}%
  \hspace{8pt}%
  \includegraphics[width=0.46\linewidth]{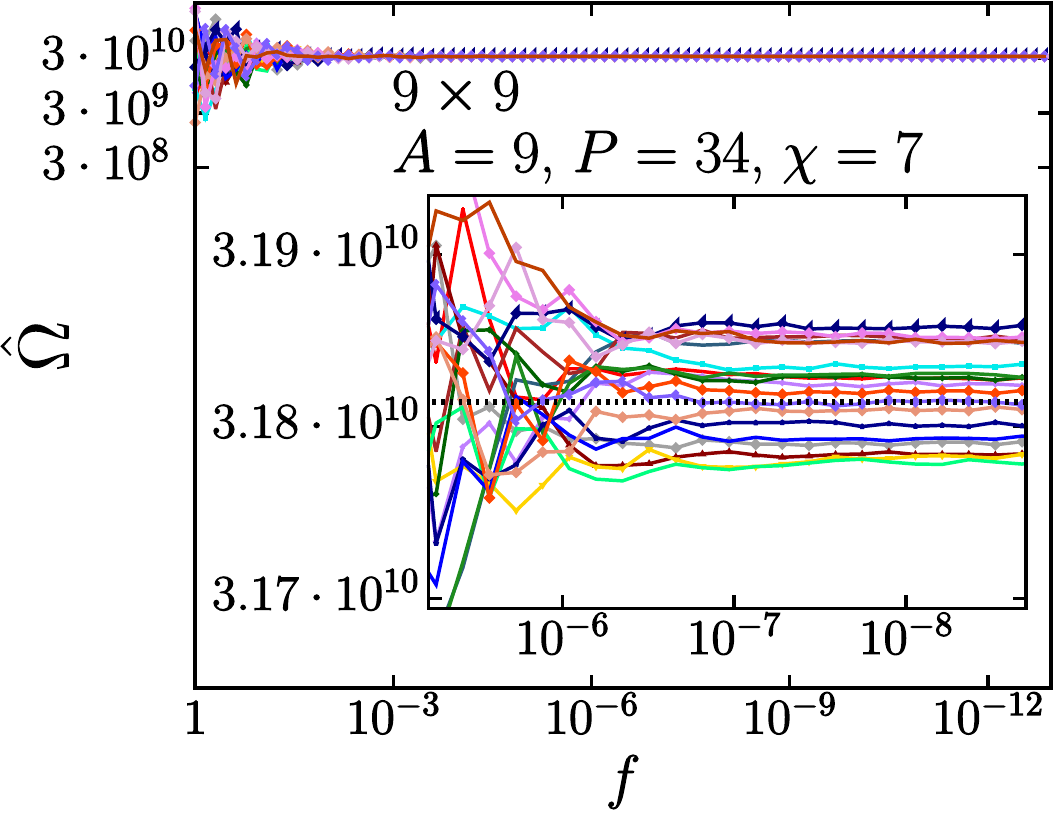}\\
  \includegraphics[width=0.46\linewidth]{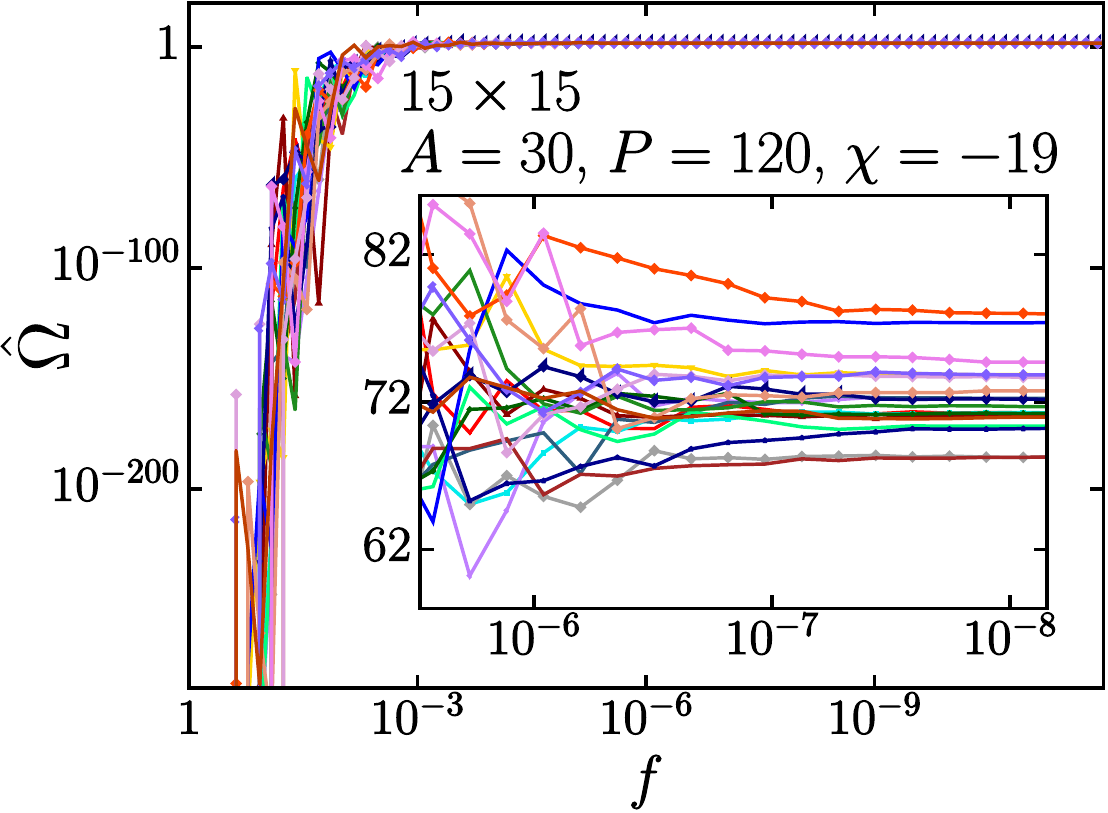}%
  \hspace{8pt}%
  \includegraphics[width=0.46\linewidth]{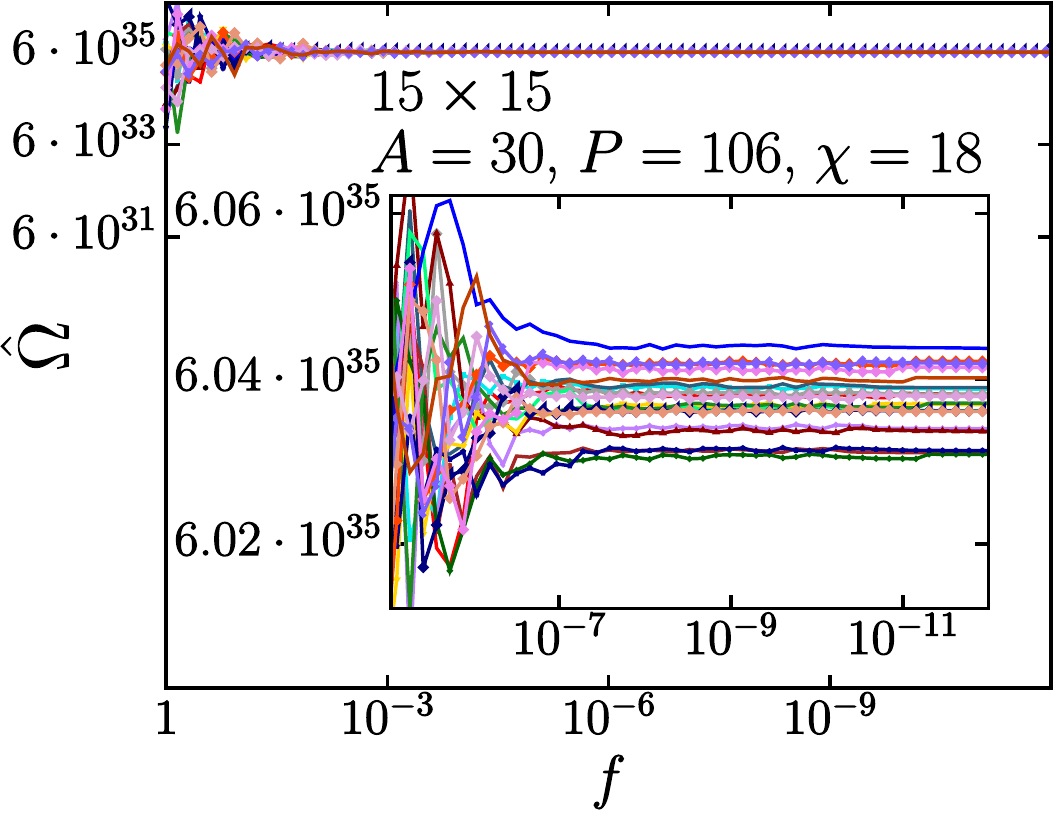}\\
  \includegraphics[width=0.46\linewidth]{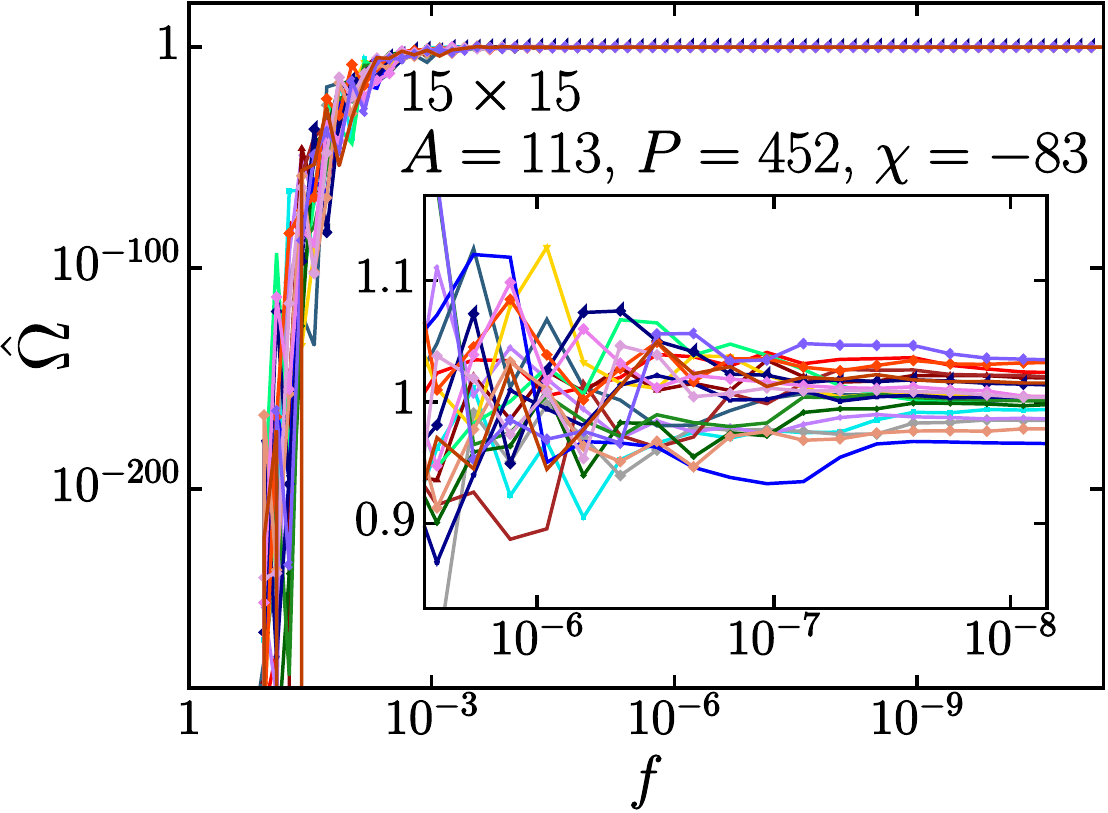}
  \hspace{8pt}%
  \includegraphics[width=0.46\linewidth]{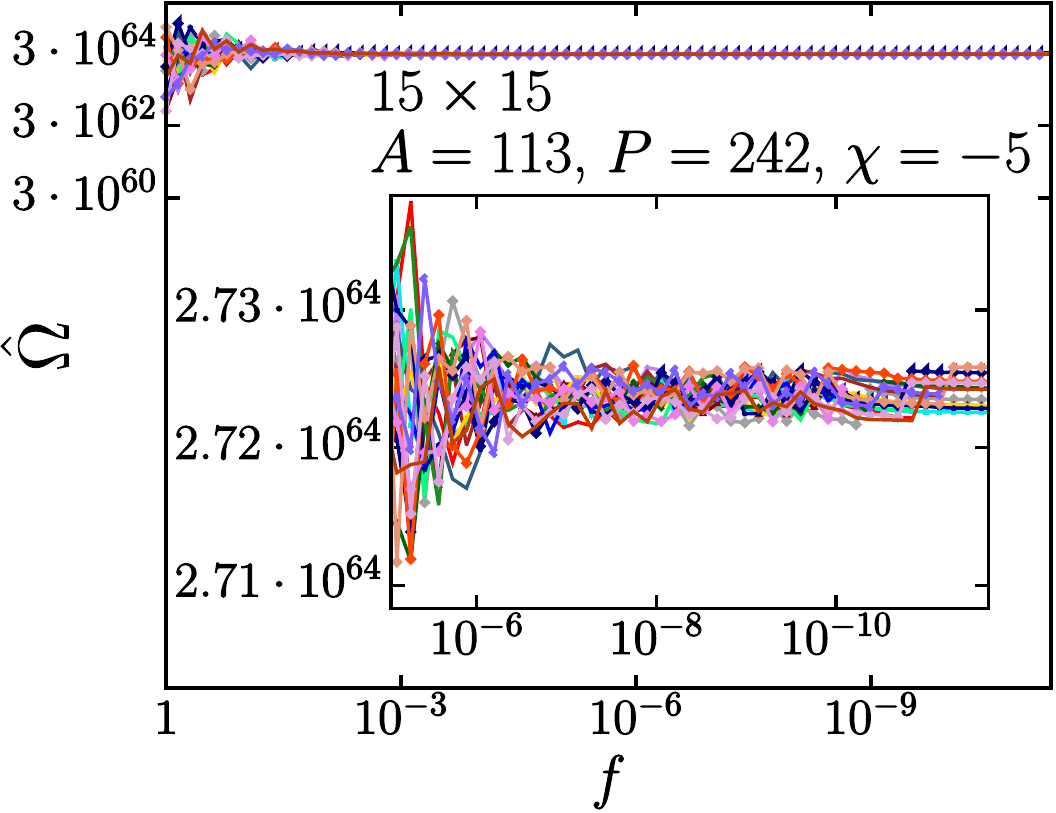}
  \caption{Saturation of error for the Wang-Landau sampling: numerical
    estimate~$\hat{\Omega}$ of the DoS as a function of the
    modification factor~$f$ that decreases when the histogram of the
    macrostates gets sufficiently flat. Each subfigure shows the
    estimates of 20 independent simulations for a different
    macrostate; these macrostates were chosen because they have
    minimum or maximum $\hat{\Omega}$ for the given window size and
    number of black pixels. The dashed lines in the top figures are the
    analytic DoS.}
  \label{fig_gamma_saturation_of_error}
\end{figure}

For the morphometric analysis, we replaced the energy by a macrostate
characterized by the area $A$, the perimeter $P$, and the Euler
characteristic $\chi$.
However, the simple implementation where each step changes the color
of a random pixel is too inefficient for finding the DoS of all three
Minkowski functionals $A$, $P$, and $\chi$. The configuration space is
too large and converges too slowly on modern hardware, e.g., for a
$15\times 15$ b/w image there are about a million macrostates, i.e.,
possible values of area, perimeter, and Euler
characteristic\footnote{Compare to the number of possible values of
  the area alone: 225.}. Moreover, the probability distribution is not
normalized, which could lead to an unknown systematic error.

We achieved the important breakthrough
by incorporating the analytically known number of configurations with
a given area $A$, the binomial coefficient $\binom{N^2}{A}$: instead
of randomly choosing a pixel and changing its color, the number of
black pixels, i.e., the area $A$, is fixed and a random black pixel is
chosen and randomly shifted to a formerly white pixel; the black
pixels perform random jumps. Note that the scanning of the energy
space is still ergodic although each simulation is restricted to a
subset because the calculations are repeated for every possible value
of the area $0\leq A\leq N^2$. Because of this separation of the
energy space into disjoint subsets, the DoS has to be determined only
w.r.t. the perimeter and the Euler characteristic. For example for a
$15\times 15$ window, the number of macrostates given the number of
black pixels remains below $10^4$, two orders of magnitude smaller
than the number of all macrostates defined by area $A$, perimeter $P$,
and Euler characteristic $\chi$. Moreover, less memory is needed, and
further optimizations are possible which further decrease the
computation time by at least one order of magnitude. Depending on the
number of black pixels, a single simulation needs from a few minutes
to about two days on a single core of Intel Xeon E3-1280 processor
(3.5 GHz). The separate simulation for different areas $A$ also allows
for a trivial and thus perfect parallelization.
This combination of numerical estimates and analytic knowledge also leads to perfectly normalized probability distributions, which follows from Eq.~\eqref{eq_gamma_PandDos}
\begin{align}
  \begin{aligned}
    \sum_{A,P,\chi}\mathcal{P}(A,P,\chi) &= \sum_{A=0}^{N^2}
    p_{\rho}^A \cdot (1-p_{\rho})^{N^2-A} \sum_{P,\chi}
    \Omega(A,P,\chi)\\
    &= \sum_{A=0}^{N^2} p_{\rho}^A \cdot (1-p_{\rho})^{N^2-A}\cdot
    \binom{N^2}{A} = 1
  \end{aligned}
  \label{eq_gamma_perfect_normalization}
\end{align}
because we use $\sum_{P,\chi} \Omega(A,P,\chi) = \binom{N^2}{A}$ as a
normalization of the DoS for a given area $A$. Moreover, for either
small or larger values of the area $A$, the DoS can be determined
analytically, which is, e.g., important for pointlike sources, which
result in a few black pixels at high thresholds.
Using this trick, we determined the DoS for scan windows up to
$20\times 20$, i.e., for systems with up to $10^{120}$
microstates. However, for this series of articles $15\times 15$ windows are
sufficient.
The method is applicable to any boundary condition. Here, the
calculations are carried out for open boundary conditions.
As described in the first paper of this series, these boundary conditions are advantageous for detecting clusters of high number of counts.

\begin{figure}
  \centering
  \includegraphics[width=0.905\linewidth]{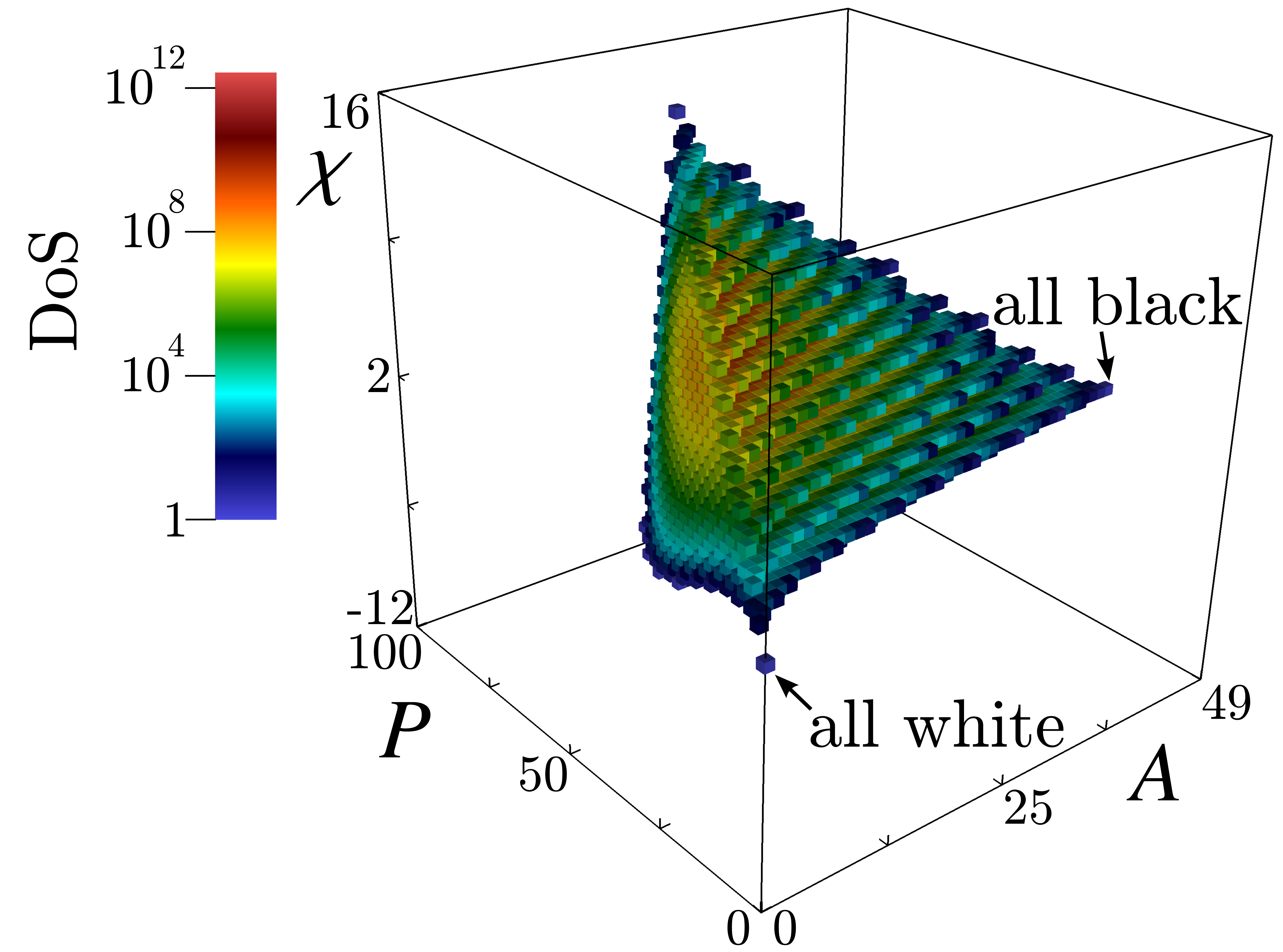}\\
  \vspace*{0.2cm}
  \includegraphics[width=0.905\linewidth]{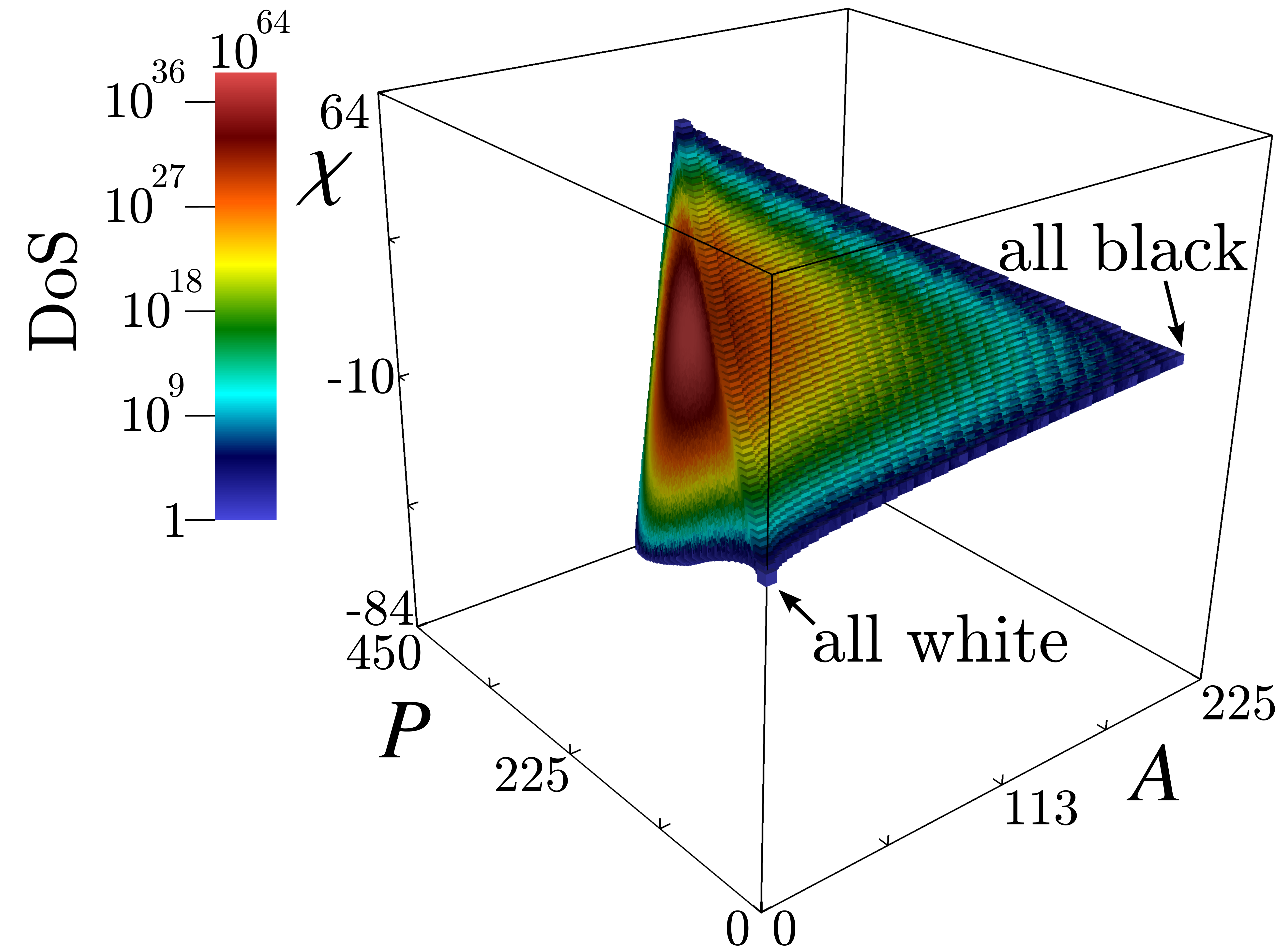}
  \caption{Density of states (DoS): the number of b/w images for a
    given area $A$, perimeter $P$, and Euler characteristic $\chi$
    with open boundary conditions: at the top, for a $7\times 7$
    window; at the bottom, for a $15\times 15$ window. For the latter,
    the color code saturates at $10^{36}$. The unit of area is a
    pixel.}
  \label{fig_gamma_DoS}
\end{figure}

\subsection{Saturation of error}

Belardinelli and Pereyra showed that the original algorithm does not
converge, but the systematic error
saturates~\citep{BelardinelliPereyra2007PRE,
  BelardinelliPereyra2007JCP, BelardinelliPereyra2008PRE}. They show
that is could be corrected for in an optimal way by choosing the
modification proportional to the inverse of time.

However, here it is sufficient to use the standard algorithm with a slight modification resulting in a slower convergence:
Instead of taking the square root if a flat histogram is reached, $f$ is only
replaced by $f^{0.7}$. With these parameters, the systematic error
then remains well below the statistical
one. Figure~\ref{fig_gamma_saturation_of_error} shows estimates
$\hat{\Omega}$ of the DoS as a function of the modification factor $f$
for several macrostates for different window sizes and number of black
pixels; in each case 20 independent estimates are shown. The insets
show the onset of the saturation of error at $f < 10^{-7}$. So, the
simulation is stopped if the modification factor drops below $10^{-7}$
before the error saturates. For the smaller systems, the outcome is
compared to the analytic result (dashed lines). Averaging over several
independent simulations leads to reliable estimates of the DoS.

The final numerical estimates of the DoS $\bar{g}$ are averaged over
eight independent simulations for each number of black pixels. The
relative statistical error of $\bar{g}$ is
$\mathcal{O}(10^{-3})$. The
compatibility is the sum of probabilities of many macrostates, see
Eq.~\eqref{eq:compatibility}; therefore, the relative error of the
deviation strength is even some orders of magnitude smaller. The trial
factor simulations in the following Section affirms the reliability of
the numerical estimates of the DoS.

\subsection{The density of states for large observation windows}

We accurately estimated the DoS of area $A$, perimeter $P$, and Euler
characteristic $\chi$ for all window sizes between $5\times 5$ to
$15\times 15$. For the smaller systems, the calculation was carried out
analytically via the above-described brute force approach, for larger
systems the DoS was estimated using the Wang-Landau method.
The cumulative computation time for all systems was about 2.3 years on a
single core of an Intel Xeon E3-1280 processor (3.5 GHz).
The densities of states are available via e-mail: michael.klatt@fau.de

Figure~\ref{fig_gamma_DoS} shows the DoS for a $7\times 7$ or a
$15\times 15$ observation window, respectively. Each voxel represents
a macrostate, and the color code plots the DoS. Note that this plot
visualizes the complete structure of a Poisson random field w.r.t. the
Minkowski functionals. The DoS reveals interesting bounds on the
possible values of area, perimeter, and Euler characteristic, which
would be interesting for a further analysis of the geometrical
properties of a Poisson random field. There are also complex features
like steps and discrete jumps between allowed macrostates; note that
these are no artifacts but appear due to the finite system size. Most
interesting for the morphometric analysis, the DoS seems to converge
fast to an asymptotic distribution that might be estimated from the
known distributions. This would allow for a global analysis of the
joint deviation strength for whole observation windows or even a
galactic scan. At least, this could be used as an initial estimate of
the DoS for the Wang-Landau algorithm which would lead to a fast
convergence even for huge system sizes.


\section{Trial correction}
\label{sec_gamma_trial_factor}

Using the exact calculations and precise estimates of the density of states from the previous section, we can now calculate for any threshold the probability distribution of the Minkowski functionals, see Eq.~\eqref{eq_gamma_PandDos}.
Thus we can determine for any measured values $(A,P,\chi)$ of the Minkowski functionals the corresponding deviation strength $\mathcal{D}(A,P,\chi)$, see Eqs.~\eqref{eq:compatibility} and \eqref{eq:dev_strength}.
Before we study the sensitivity of the morphometric analysis and the advantage of the additional structure characterization in the third paper of this series,
we first have to confirm that there is no overestimation of the significance due to repeated trials.

As described in Sec.~\ref{sec_gamma_short_overview}, we determine the deviation strength at each threshold and compare the maximum over all thresholds to the null hypothesis criterion.
These repeated (though dependent) trials increase the likelihood of a statistically significant fluctuation in the background noise.
When we correct for this trial effect, we have to check whether a pronounced difference appears for the deviation strength $\mathcal{D}(A)$ based on a single Minkowski functional or for $\mathcal{D}(A,P,\chi)$ based on all three functionals.
We show that this is not the case.
The increase in sensitivity, which we will present in the next paper in this series, is not a result of an increased effective number of trials, but of the additional structure information.

\subsection{A conservative yet efficient trial correction}

\begin{figure}[t]
  \centering
  \subfigure[][]{%
    \includegraphics[width=\linewidth]{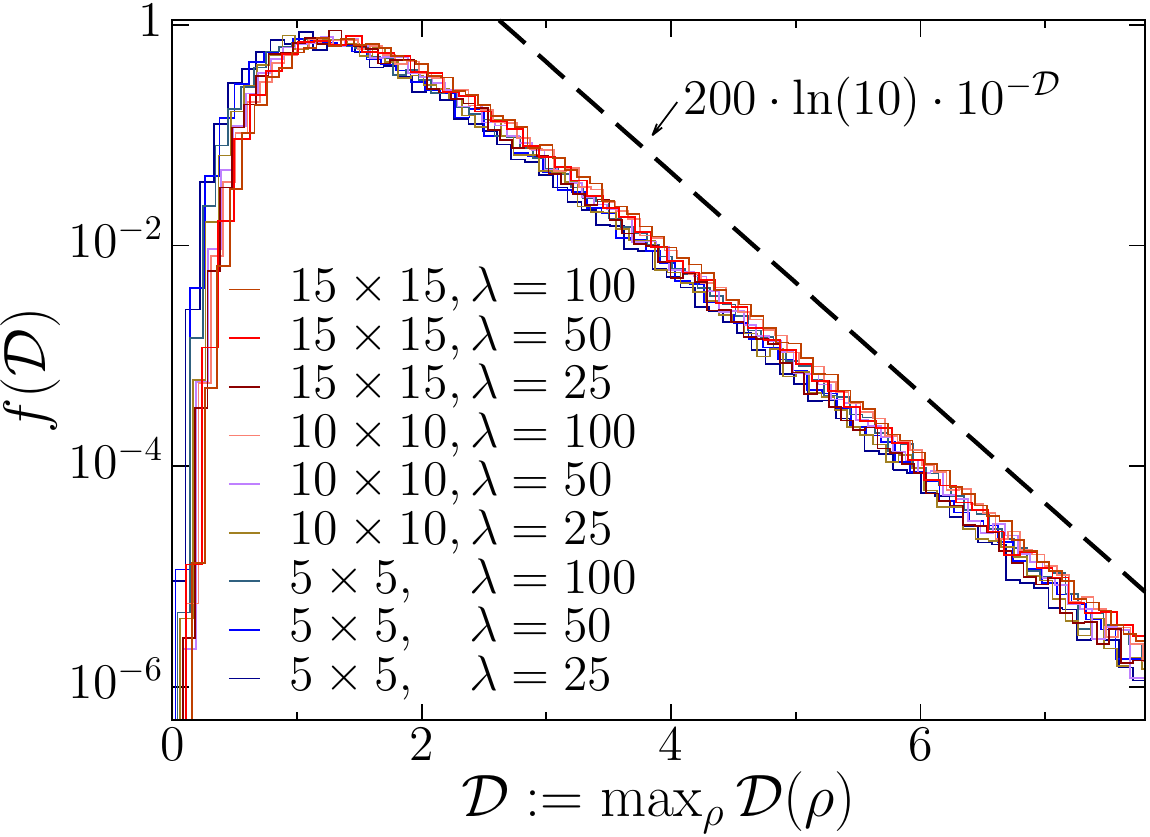}
    \label{fig_gamma_trial_a}
  }\\
  \subfigure[][]{%
    \includegraphics[width=\linewidth]{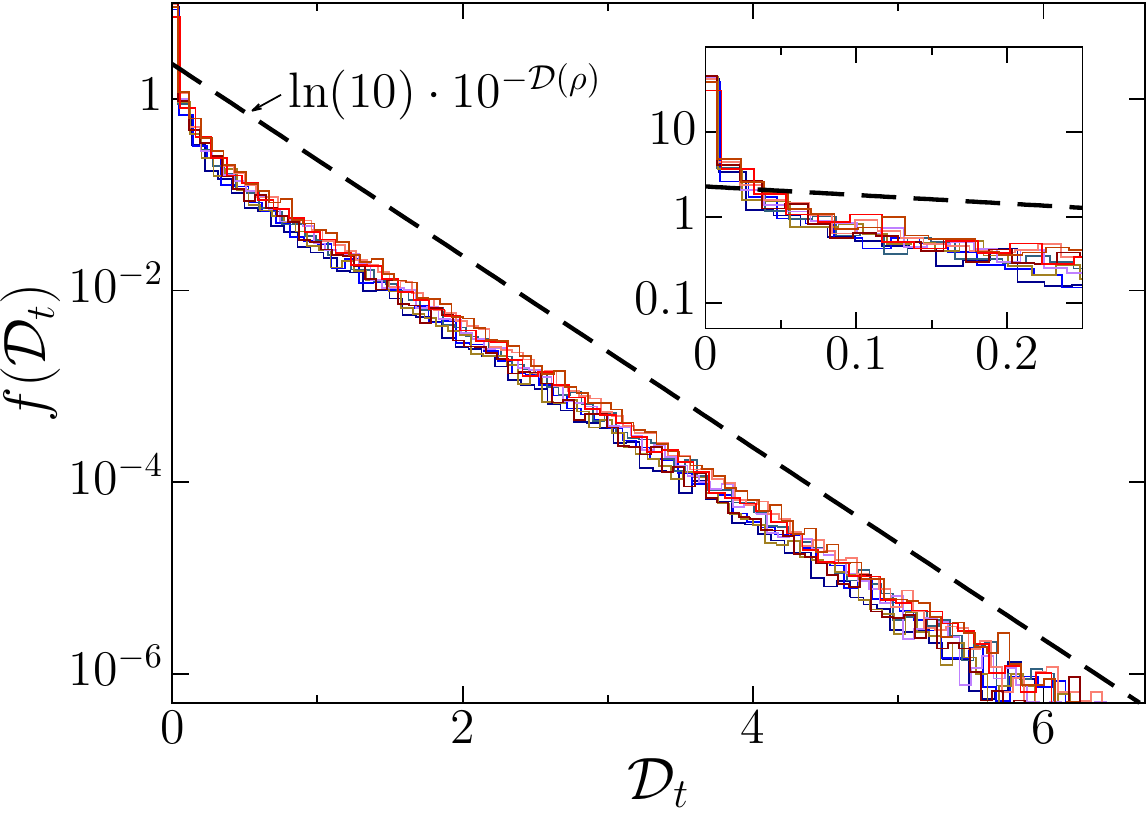}
    \label{fig_gamma_trial_b}
  }%
  \caption{Trial correction: (a) Empirical Probability Density
    Function (EPDF) of the maximum of the deviation strength for pure
    Poisson background signals at different intensities $\lambda$ and
    for different sizes of the observation windows. The dashed line shows an
    upper bound for the distribution of the deviation strength; for
    large $\mathcal{D}$ this corresponds to the distribution of the
    maximum of 200 trials, which is in this series of papers an upper bound on
    the number of thresholds; (b) EPDF of the trial corrected
    deviation strength $\mathcal{D}_t$ using
    Eq.~\eqref{eq_gamma_postTrial} with $n_t=200$. The dashed line
    shows the probability density function of the deviation strength
    $\mathcal{D}(\rho)$ of a single trial at threshold $\rho$. Inset:
    close up of the peak at $\mathcal{D}_t\rightarrow 0$.}
  \label{fig_gamma_trial}
\end{figure}

Taking the maximum of the deviation
strengths over all thresholds corresponds to repeated null hypothesis tests for
the different b/w images. Each b/w image can be seen as a single
trial, but the more trials, the higher the probability for a
significant random fluctuation. Because the different b/w images are
constructed from the same count map, they are obviously closely
correlated, which reduces the trial effect. Deriving the exact
distribution of maximum deviation strengths is, however, not
analytically solvable. A $15\times 15$ count map with not more than
$100$ counts in each bin has $10^{450}$ possible configurations.

By assuming that the different b/w images are independent of
each other, a conservative estimate of the trial correction was suggested in the first paper of this series:
given $n_t$ independent trials, the probability to find a
compatibility lower than $\alpha$ is for each single b/w image
$\alpha$; the probability to find no such deviation in any of the
$n_t$ independent trials is $(1-\alpha)^{n_t}$. Thus, the probability
to find at least one trial with a compatibility lower than $\alpha$ is
given by
\begin{equation}
  \alpha_{n_t} = 1-(1-\alpha)^{n_t} \, .
\end{equation}
For $\alpha \ll 1:\ \alpha_{n_t} = n_t \alpha +
\mathcal{O}(\alpha^2)$, i.e., the influence of the repeated trials can
simply be taken into account by multiplying the compatibility with a
constant, the so-called trial factor $n_t$. The deviation strength is
corrected by
\begin{equation}
  \mathcal{D}_{n_t} = -\log_{10}\left(1-(1-10^{-\mathcal{D}})^{n_t}\right),
  \label{eq_gamma_postTrial}
\end{equation}
which is for a large deviation strength ($\mathcal{D}>3$) well
approximated by a constant offset: $\mathcal{D}_{n_t} \approx
\mathcal{D} - \log_{10}(n_t)$.

Setting $n_t$ equal to the number of different thresholds provides a
simple and conservative estimate. Figure~\ref{fig_gamma_trial_a}
shows the Empirical Probability Density Function (EPDF) of the maximum
of the deviation strength for different intensities and differently
large observation windows derived from $10^8$ Poisson fields for each
system. Even a constant upper bound on the number of thresholds
$n_t=200$ provides a close upper bound for the trial correction.
Therefore, in this series of papers with intensities well below $\lambda=200$,
the maximum of the deviation strength $\mathcal{D}$ over all
thresholds is corrected for this trial effect by mapping to
$\mathcal{D}_{n_t}=-\log_{10}\left(1-(1-10^{-\mathcal{D}})^{n_t}\right)$
in Eq.~\eqref{eq_gamma_postTrial} with $n_t=200$. The EPDF of this
trial corrected deviation strength is plotted in
Fig.~\ref{fig_gamma_trial_b}. It is compared to the probability
density function of the deviation strength $\mathcal{D}(\rho)$ of a
single trial at threshold $\rho$, i.e., the actual claim of
significance. Only in the case of nearly perfect agreement with the
background structure, i.e., for $\mathcal{D}_t < 0.05$ (see inset),
the empirical distribution has a strong peak. For all deviation
strengths of interest, the estimate of $\mathcal{D}_t$ is conservative
yet relatively close to the optimum (dashed line).

Note that the fluctuations of the distribution at relatively small
deviation strengths are no statistical fluctuations but are
reproducible features of the actual distribution which is highly
irregular because of the finite system size.

\subsection{Joint and simple deviation strength}

If the deviation from the background structure is quantified either only by the area or by all three Minkowski functionals, the trial factor might change.
The effective number of trials might change due to the differently detailed characterization of the single black and white images.

\begin{figure}
  \centering
  \includegraphics[width=\linewidth]{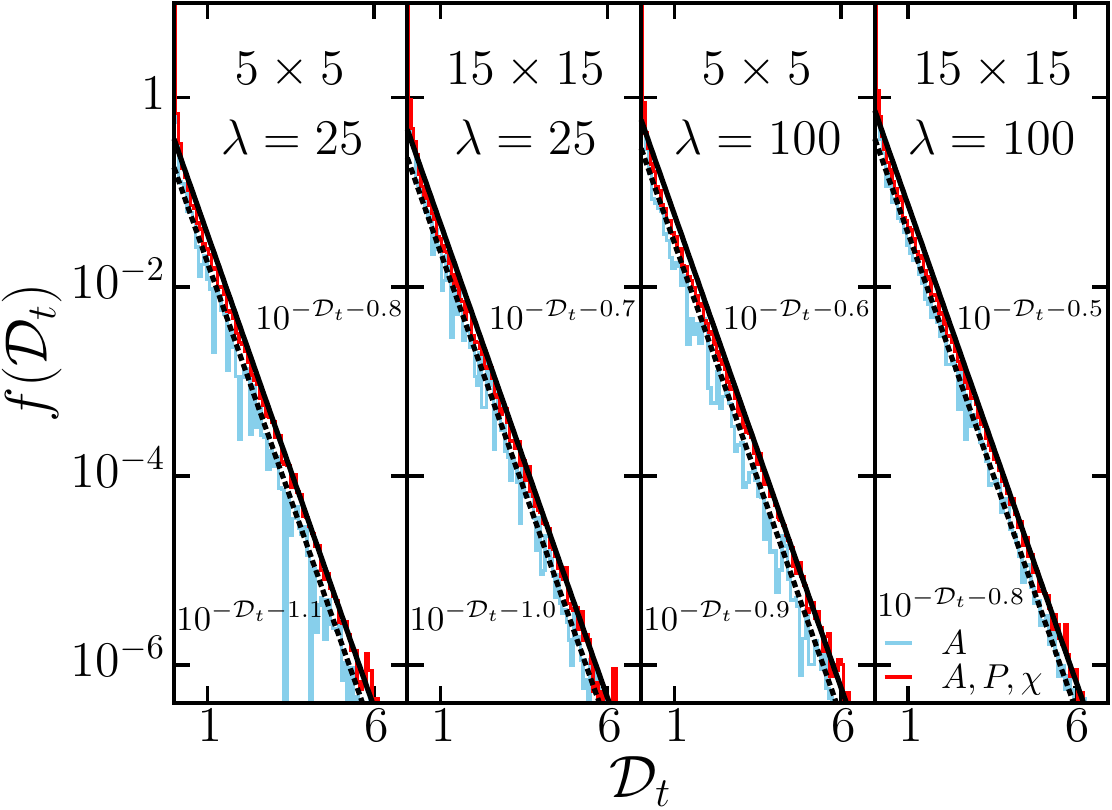}
  \caption{Comparison of the simple and joint deviation strength:
    empirical probability density $f$ of the trial corrected
    deviation strength $\mathcal{D}_t$ w.r.t. only the area ($A$) or all
    three Minkowski functionals ($A,P,\chi$), respectively.
    To detect possible deviations between the simple and joint deviation strength, the
    distributions are approximated by shifted exponential
    distributions $\ln(10)10^{-\mathcal{D}_t-c}$ for
    different system sizes and background intensities $\lambda$.}
  \label{fig_gamma_trial_A_vs_APC}
\end{figure}

In order to compare the joint deviation strength, i.e., w.r.t. all
Minkowski functionals, to the simple deviation strength w.r.t. only
the area, Figure~\ref{fig_gamma_trial_A_vs_APC}
shows both the distributions of the trial-corrected simple and joint
deviation strengths.
The binned distribution of the simple deviation strength, as depicted
in Fig.~\ref{fig_gamma_trial_A_vs_APC}, is highly irregular, even more irregular
than the distribution of the joint deviation strength.
These are again no statistical fluctuations, but reproducible features
of the complicated distribution function.
The strong irregularity for the simple deviation strength arises because of the small number of macrostates.

To detect any systematic difference that might appear, we approximate the distributions (for different system sizes and background intensities)
by shifted exponential distributions $\ln(10)10^{-\mathcal{D}_t-c}$.
The parameter $c=\mathcal{O}(1)$ measures how conservative the above described trial correction is.
The maximum difference in $c$ that we can find for the simple and for joint deviation strength is only about $0.3$.
This value is negligible for the comparison between joint and the simple deviation strength in the third paper of this series.
The increase in the deviation strength due to the joint structure characterization of all Minkowski functionals will be much larger.

For convenience and because all results in the third paper of this series are trial corrected,
we omit the subscript $t$, which indicates the trial correction, i.e., $\mathcal{D}$ is used from now on instead of $\mathcal{D}_t$.


\section{Conclusion}
\label{sec_conclusion}

The morphometric analysis using Minkowski functionals allows to detect sources in VHE gamma-ray sky maps via structural deviations from the background noise.
However, it does not require any prior knowledge about the source, because it characterizes the shape of the background noise itself.

Therefore, we derive an accurate descriptor of the complex structure of these background events in gamma-ray astronomy by combining techniques from stochastic geometry and statistical physics.
The main difficulty is to gain a detailed knowledge about the background fluctuations, in other words, a precise estimate of the joint distribution of area, perimeter, and Euler characteristic.
This problem can easily be reformulated in finding the density of states, i.e., the number of b/w images with given values of the Minkowski functionals, see Eq.~\eqref{eq_gamma_PandDos}.
However, the total number of configurations of a $15\times 15$ b/w image is $\mathcal{O}(10^{67})$ and the density of states for different macrostates can vary from $\mathcal{O}(10^{1})$ to $\mathcal{O}(10^{64})$.
To derive accurate estimates of the density of states even for large observation windows up to $15\times 15$, we combine analytic knowledge of the structure distributions with a very efficient algorithm
from statistical physics for estimating density of states, the so-called Wang-Landau algorithm, see Fig.~\ref{fig_gamma_saturation_of_error}.

We explicity derive even for observation windows larger than $7\times 7$ accurate estimates of the distribution of the background structure, which is simultaneously characterized by all Minkowski functionals, see Fig.~\ref{fig_gamma_DoS}.

Such a refined structure characterization can extract more information out of the same data and formerly undetected sources can eventually be detected.
We study this increase in sensitivity in the third paper of this series.

Our approach is here used to derive an accurate estimate of the structure of a discrete Poisson random field quantified by Minkowski functionals.
Note, however, that it can in principle be used for even more general random black-and-white pixelated images and methods of structure characterization to efficiently calculate their probability distributions.

Because of the repeated null hypothesis tests at different thresholds, a trial correction is needed.
Here we show in a simulation study that assuming independent trials at different thresholds provides a conservative yet close estimate for trial correction, see Fig.~\ref{fig_gamma_trial}.
Moreover, we present a local detector acceptance correction in Appendix~\ref{sec_gamma_detector_acceptance}.
We demonstrate that it increases the sensitivity of the morphometric analysis in the application to real data significantly, see Fig.~\ref{fig_gamma_local_detector_acceptance}.

The here developed techniques allow to determine the DoS of sliding windows of up to about $20\times20$ pixels.
If even larger window sizes were necessary, an initial guess of the corresponding DoS could be achieved by a rescaling of the number of configurations for windows with $20\times20$ pixels.
The initial guess could then lead to a faster convergence of the Wang-Landau algorithm.
For arbitrarily large scan-windows, the expectations and the covariances of the Minkowski functionals are known analytically,
which can be used for an alternative test statistic based on empirical cumulative distribution functions; see~\citet{Klatt2016}.

\begin{acknowledgements}
  We thank Christian Stegmann and Daniel G\"oring for valuable discussions, suggestions, and advice.
  We thank the German Research Foundation (DFG) for the Grant No. ME1361/11 awarded as part of the DFG-Forschergruppe FOR 1548 ``Geometry and Physics of Spatial Random Systems.''
\end{acknowledgements}

\appendix

\section{Detector acceptance correction}
\label{sec_gamma_detector_acceptance}

Observations in gamma-ray astronomy are affected by a spatially varying detector acceptance, that is, for each bin~$i$ only a fraction $f_i$ of the signals are expected to be detected.
If a model of the camera acceptance is available, we can take these into account when analyzing real data.

In the first paper of this series, we presented a correction that regained an isotropic and homogeneous structure for background measurements.
It is shortly repeated in Sec.~\ref{sec:MCobserv} and complemented in Sec.~\ref{sec_gamma_post_selection} by the possibility to subtract point sources.
Both techniques are combined in Sec.~\ref{sec_gamma_combined_detector} for a slightly improved analysis.
In Sec.~\ref{sec_gamma_local_detector}, we extend this concept to an optimal correction of the detector acceptance.
It locally adjusts the analysis to the variations in the acceptance, resulting in a minimal suppression of source signals.

\subsection{Monte Carlo observations}
\label{sec:MCobserv}

A simple weighting with $1/f_i$ (to correct the reduced acceptance rate $f_i < 1$ in bin $i$) would destroy the Poisson structure of the sky map, because fractional photon counts could occur.
Instead we use the null hypothesis that the background noise is a homogeneous Poisson process with intensity~$\lambda$.
Then the detected number of counts in bin $i$ is a Poisson random variable with mean value $\lambda_i = f_i \cdot \lambda$.
By adding a new Poisson-distributed random variable (a Monte Carlo observation) with mean value~$\lambda_i^+ = (1-f_i) \cdot \lambda$,
the original random process with mean value~$\lambda$ is regained.

Because the pseudo-photon counts from the Monte Carlo observations fulfill the null hypothesis by construction,
they cannot introduce additional structural deviations from the homogeneous isotropic Poisson field. 
Moreover, if $f_i \approx 1$, e.g., in the center of the field of view, the number of counts is nearly unchanged.
However, source features in regions where $f_i \ll 1$ are covered by the additionally created pseudo-photon counts.

\subsection{Postselection}
\label{sec_gamma_post_selection}

If there is a strong pointlike source within the field of view, it
needs to be subtracted from the data in order to detect diffuse
gamma-ray signals, i.e., broad sources in the same field of
view~\citep{galcen}.  For the morphometric analysis, a naive
multiplication with a factor is not appropriate, because instead of a
homogeneous Poisson field this would again produce fractal
counts.

Instead we here present a postselection, which is based on a new null hypothesis that there is
no additional source, i.e., that there is only a uniform Poisson noise
with intensity $\lambda$ in each bin and the predicted pointlike
source with an intensity $\Lambda_i$ in bin $i$. On average, the
number of counts must be reduced by a factor of
$\lambda/(\lambda+\Lambda_i)$. However, the reduction must maintain
the integer number of counts.

This is achieved by a Monte Carlo postselection of the signals: only $n$ of the $k$ measured signals is
kept, where $n$ follows a Binomial distribution with probability
$\lambda/(\lambda+\Lambda_i)$. 
The resulting random field is a homogeneous Poisson field with
intensity $\lambda$ where the estimate is conservative and thus stable
against errors in the prediction of the model of the pointlike source, see \citet{Klatt2010}.

\subsection{Combined postselection and MC observations} 
\label{sec_gamma_combined_detector}

The detector acceptance correction by adding Monte Carlo signals to
gain a homogeneous Poisson field can be very conservative, especially in
regions of very low acceptance $f_i\ll 1$. An additional Poisson random
number suppresses the detection of a significant excess or structural
deviation in the actual counts.

To improve upon this, \citet{Goering2012} combined the Monte
Carlo observations and the postselection: the signals are not on
average filled up to the original expected intensity $\lambda$, but to
a chosen level $f\cdot\lambda$, where $f$ is in-between the maximum
and minimum acceptance. For bins with $f_i < f$ MC observations are
added to the data set, on average $(f - f_i)\lambda$ events are added,
and for bins with $f_i > f$ the signals are reduced via the
postselection as described above, on average $f/f_i$ signals are
kept. By setting $f$ to the corresponding value $f_j$ in the region of
interest, a nearly optimal detector acceptance correction can be
achieved for a limited region with $f_i \approx f_j$. However, for the
rest of the observation window the estimate is still very conservative.

\subsection{Local detector acceptance correction}
\label{sec_gamma_local_detector}

\begin{figure}
  \centering
  \subfigure[][]{%
    \includegraphics[width=0.32\linewidth]{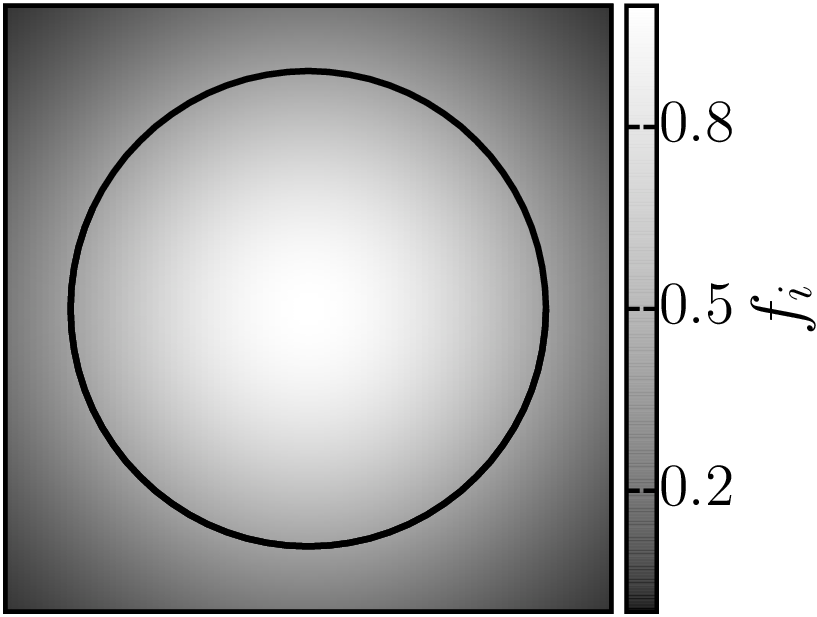}%
    \label{fig_gamma_detector_1}%
  }%
  \hspace{4pt}%
  \subfigure[][]{%
    \includegraphics[width=0.32\linewidth]{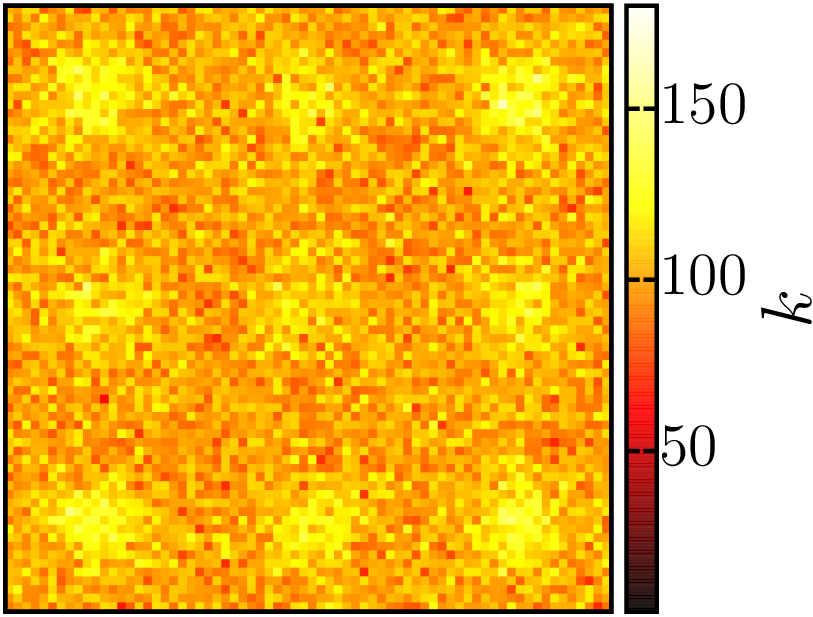}%
    \label{fig_gamma_detector_2}%
  }%
  \hspace{4pt}%
  \subfigure[][]{%
    \includegraphics[width=0.32\linewidth]{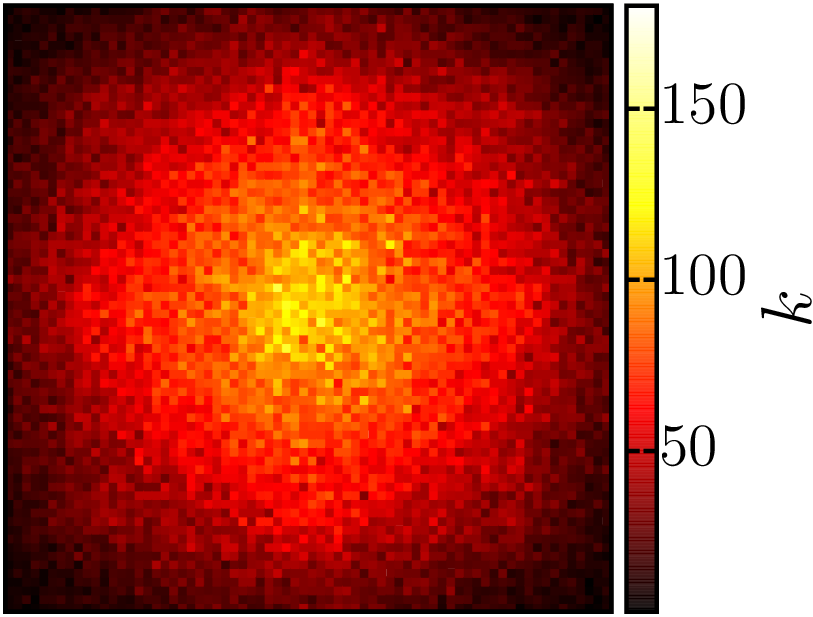}%
    \label{fig_gamma_detector_3}%
  }\\
  \subfigure[][]{%
    \includegraphics[width=0.32\linewidth]{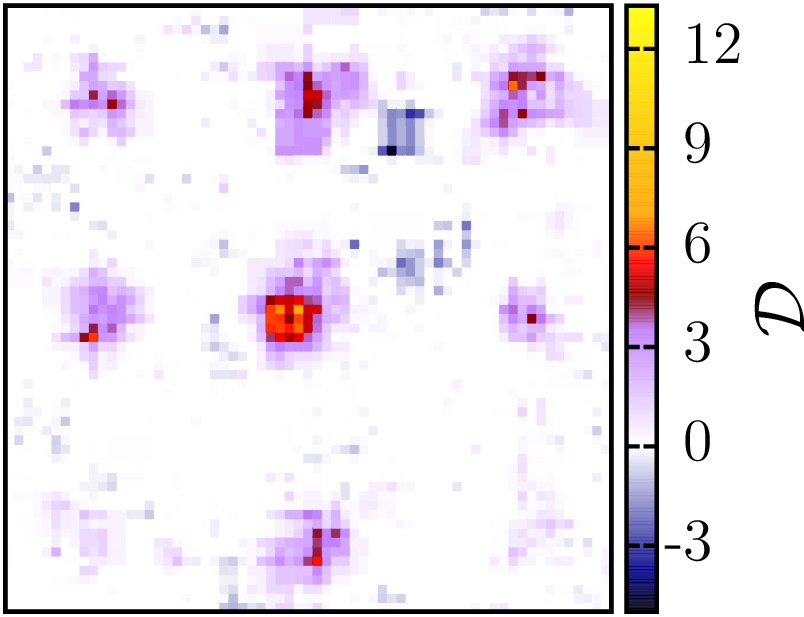}%
    \label{fig_gamma_detector_4}%
  }%
  \hspace{4pt}%
  \subfigure[][]{%
    \includegraphics[width=0.32\linewidth]{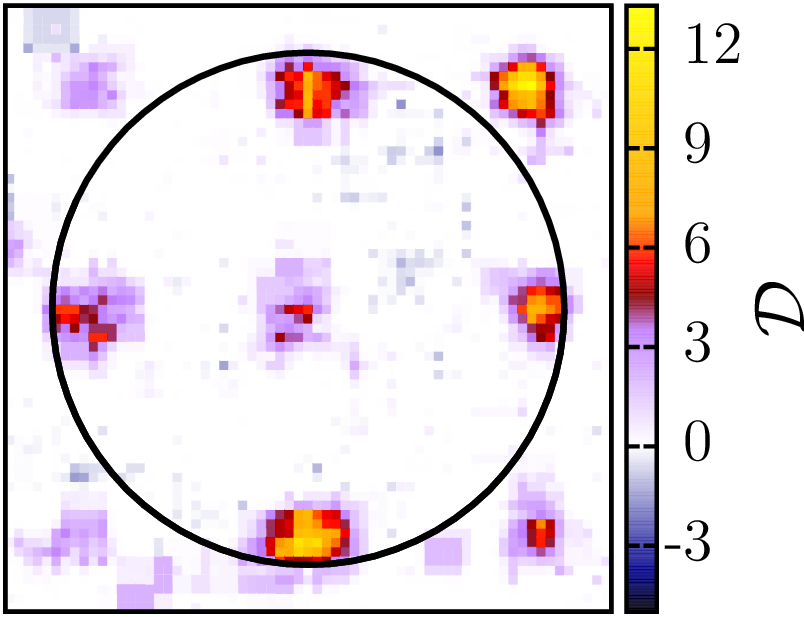}%
    \label{fig_gamma_detector_5}%
  }%
  \hspace{4pt}%
  \subfigure[][]{%
    \includegraphics[width=0.32\linewidth]{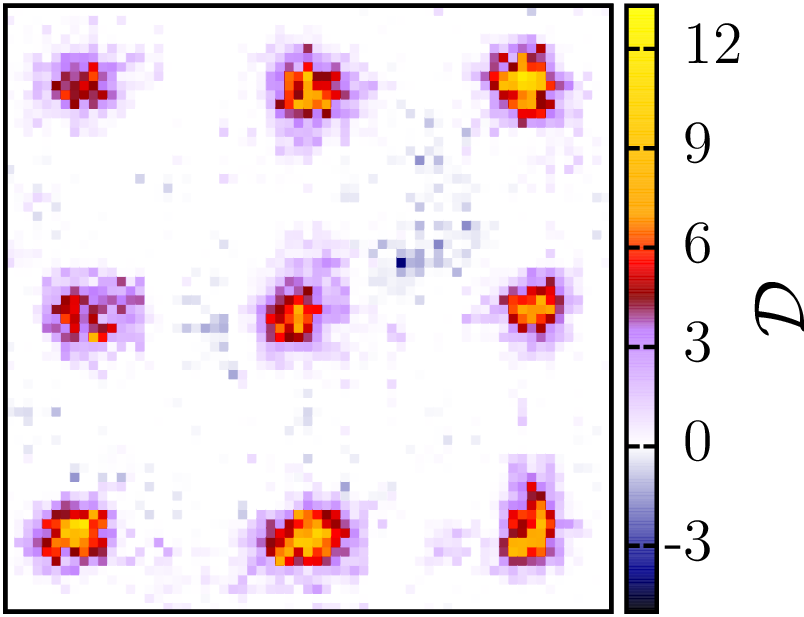}%
    \label{fig_gamma_detector_6}%
  }\\
  \subfigure[][]{%
    \includegraphics[width=0.32\linewidth]{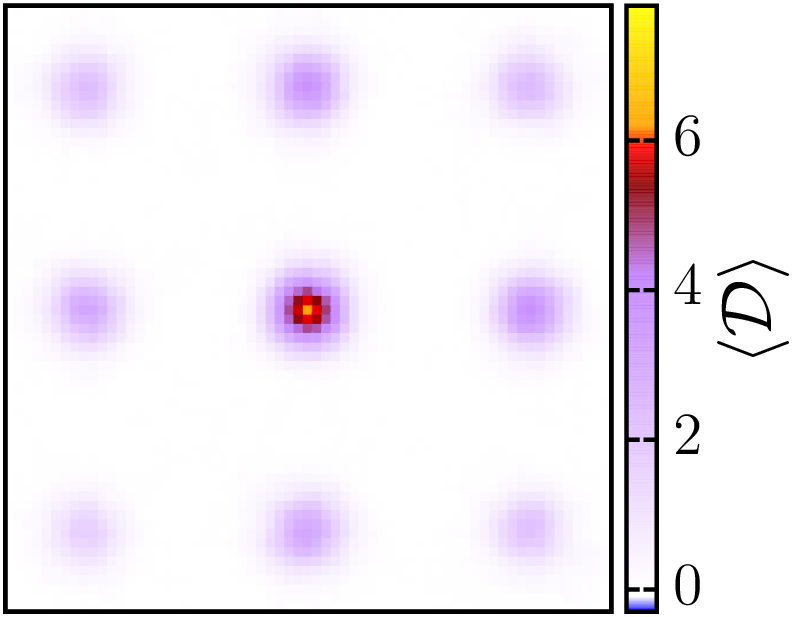}%
    \label{fig_gamma_detector_7}%
  }%
  \hspace{4pt}%
  \subfigure[][]{%
    \includegraphics[width=0.32\linewidth]{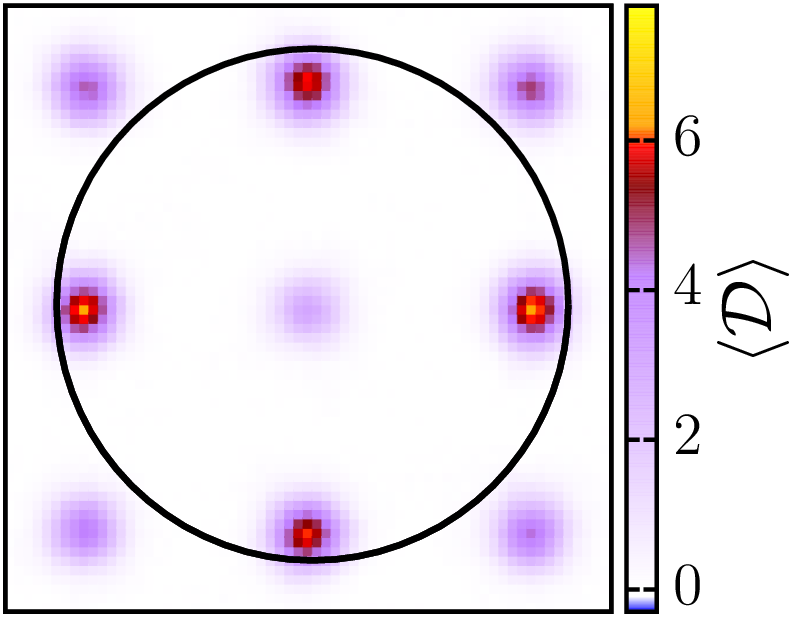}%
    \label{fig_gamma_detector_8}%
  }%
  \hspace{4pt}%
  \subfigure[][]{%
    \includegraphics[width=0.32\linewidth]{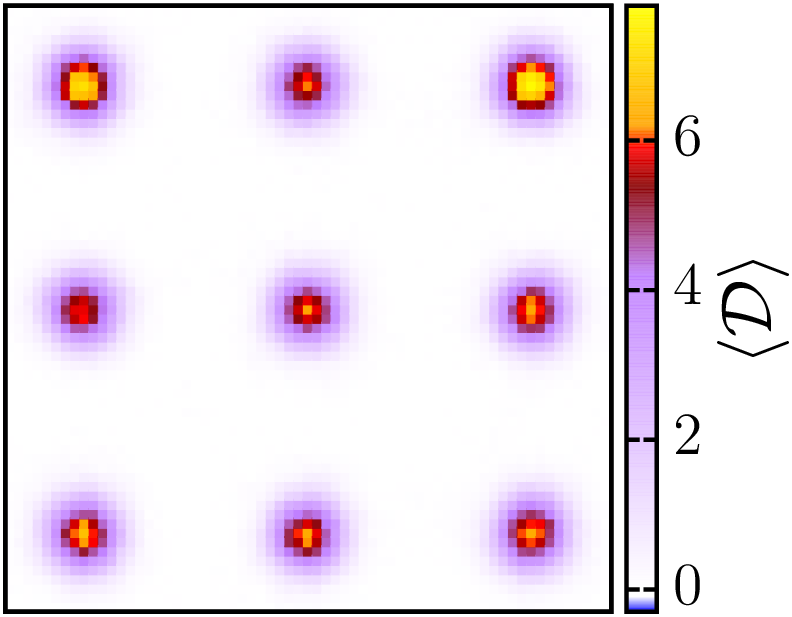}%
    \label{fig_gamma_detector_9}%
  }
  \caption{Detector acceptance correction: (a) acceptance map $f_i$;
    (b) unreduced simulated count map; (c) counts are randomly reduced
    according to acceptance map; (d--f) Minkowski sky maps analyzing
    the count map depicted in (c) for which the acceptance is
    corrected (d) globally via Poisson filling, (e) globally via both
    Poisson filling and postselection with the target intensity
    $\lambda/2$ (the corresponding region is indicated by a black
    circle), and (f) locally; (g--i) for each pixel an average over
    100 Minkowski sky maps where the acceptance is corrected (g)
    globally via Poisson filling, (h) globally via both Poisson
    filling and postselection with the target intensity $f_i=0.5$, and
    (i) locally. Only if the acceptance is corrected locally, can in
    this example a single Minkowski sky map detect all sources at any
    distance to the center.}
  \label{fig_gamma_local_detector_acceptance}
\end{figure}

The sensitivity of the morphometric analysis can be tremendously
increased by changing from a global detector acceptance as described
above to a local detector acceptance. Instead of correcting the
detector acceptance globally and prior to the construction of the
Minkowski sky map, it is performed separately and independently for each
position of the sliding window. 

Given the acceptance $f_i$ of the central bin of a sliding window, the target intensity is chosen to be
$f_i\cdot\lambda$ and the original observed counts map is corrected by
the combined postselection and MC observation. Thus, for each sliding
window an optimal acceptance correction is performed. Additionally,
this slightly decreases the correlation between the deviation strength
of overlapping observation windows, because the random variables of
the postselection and MC observation, i.e., the added or subtracted
counts, are different.

The huge gain in sensitivity is demonstrated for simulated data in
Fig.~\ref{fig_gamma_local_detector_acceptance}.
The detector acceptance is approximated in this simulation by a
Gaussian with $f_i = 1$ in the center of the sky map and the line of
half-maximum indicated as a black line in (a). First, a count map is
simulated at a homogeneous Poisson background with intensity 100 and
with nine differently strong sources at different distances to the
center, see Fig.~\ref{fig_gamma_local_detector_acceptance}~(b). Then,
for each pixel only a fraction $f_i$ of the counts are accepted
according to (a). The resulting count map is depicted in (c).
Figures~\ref{fig_gamma_local_detector_acceptance}~(d--f) are Minkowski
sky maps based on the area, the perimeter, and the Euler
characteristic analyzing the count map in figure~(c). The sliding
window size is $6\times 6$. Because the reduction and Poisson filling
at very low intensities can strongly fluctuate, 100 Minkowski sky maps
are averaged in each pixel in
Figs.~\ref{fig_gamma_local_detector_acceptance}~(g--i).

If the whole count map is corrected for a homogeneous Poisson field
with background intensity $\lambda$ by filling with Poisson counts,
see Figs.~\ref{fig_gamma_local_detector_acceptance}~(d,g), only the
central source is detected. The other source signals are overwhelmed
by the additional simulated Poisson signals. If a target intensity
$\lambda'= \lambda/2$ is chosen and a homogeneous Poisson field is
regained under the null hypothesis via both Monte Carlo observations
and postselection, see
Figs.~\ref{fig_gamma_local_detector_acceptance}~(e,h), only the
sources in the region of the target intensity can be detected; if
there is slight offset, the sources are hardly detected. The sources
in the corners can only be detected in single samples where the source
signals can fluctuate strongly. Also the source in the center,
where there is originally a perfect detector acceptance, is no longer
detected.
However, if the detector effects are corrected locally, see
Figs.~\ref{fig_gamma_local_detector_acceptance}~(f,i), the improved
background correction allows to detect all sources at any distance to
the center in a single Minkowski sky map.

Because for each pixel in the Minkowski sky map the hypothesis test is
performed for a different background intensity, the probability that a
pixel is black at a given threshold varies for all sliding
windows. Therefore, the probability distribution of the Minkowski
functionals has to be determined separately for each sliding
window. Therefore, the computation time strongly increases compared to
the former detector acceptance corrections with a global background
intensity. However, if there are strong detector effects and the
analysis is supposed to detect more than just an expected pointlike source
at a known position, the local detector correction introduced here can
provide a tremendous increase in sensitivity that justifies this
additional computational effort.

\bibliographystyle{aa}
\bibliography{gamma-II}

\end{document}